\documentclass[pra,aps,preprint,showpacs]{revtex4}

\usepackage[latin1]{inputenc}
\usepackage[intlimits,tbtags]{amsmath}
\usepackage{color}
\usepackage{amsfonts}
\usepackage{amssymb}
\usepackage{graphicx}%[draft]
\usepackage{bm}

\usepackage{color}

\newcommand{\upt}{\ensuremath{U_{\text{PT}}}}
\newcommand{\elaser}{\ensuremath{E_{\text{L}}}}

\newcommand{\bracket}[2]{\ensuremath{\langle #1|#2\rangle}}

\newcommand{\ilm}{Institut Lumi\`ere Mati\`ere, UMR5306 Universit\'e Lyon 1-CNRS, Universit\'e de
Lyon, F-69622 Villeurbanne Cedex, France}

\newcommand{\halle}{Institut f\"ur Physik, Martin-Luther-Universit\"at
  Halle-Wittenberg, D-06099 Halle, Germany}

\begin{document}
\title{Optimization of the ionization time of an atom with tailored laser pulses: a theoretical study}

\author{David Kammerlander}\email{kammerlander.david@gmail.com}
\affiliation{\ilm}
\author{Alberto Castro}
\affiliation{
ARAID Foundation, Edificio CEEI, Mar{\'{\i}}a Luna 1, 50018 Zaragoza (Spain)
}
\affiliation{
Institute for Biocomputation and Physics of Complex Systems of the University
of Zaragoza, Mariano Esquillor s/n, 50018 Zaragoza (Spain)
}
\author{Miguel A. L. Marques}
\affiliation{\ilm}
\affiliation{\halle}

\date{\today}

\begin{abstract}
How fast can a laser pulse ionize an atom? We address this
  question by considering pulses that carry a fixed time-integrated energy
  per-area, and finding those that achieve the double requirement of
  maximizing the ionization that they induce, while having the shortest
  duration. We formulate this double-objective quantum optimal control problem
  by making use of the Pareto approach to multi-objetive optimization, and the
  differential evolution genetic algorithm. The goal is to find out how much a
  precise time-profiling of ultra-fast, large-bandwidth pulses may speed up the
  ionization process with respect to simple-shape pulses. We work on a simple
  one-dimensional model of hydrogen-like atoms (the P\"oschl-Teller potential),
  that allows to tune the number of bound states that play a role in the
  ionization dynamics. We show how the detailed shape of the pulse accelerates
  the ionization process, and how the presence or absence of bound states
  influences the velocity of the process.
%Closely related to the ionization regime is the ionization time, the time it takes to ionize one atom.
%While the role of laser power is evident both in the transition between these
%two regimes and in the ionization time, the effect of the presence of another bound state is less obvious. 
%We cast the question about the minimum ionization time of a model one-electron system with a fixed amount of
%laser energy into a Pareto problem. 
%Thus we do not have only access to one optimum, i.e. the minimum time for complete
%ionization, but to all Pareto optimal solutions. In this work
%we present a physical interpretation of the shape of the line along which these optimal solutions are found.
%Furthermore we show the decisive and ambiguous role of a second bound state during ionization. 
\end{abstract}

% PACS
% 32.80.Qk  Coherent control of atomic interactions with photons 
% 32.80.Fb  Photoionization of atoms and ions
% 32.80.Rm  Multiphoton ionization and excitation to highly excited states
% 42.65.Re  Ultrafast processes; optical pulse generation and pulse compression

\maketitle

%----------------------------------------
% INTRODUCTION
%----------------------------------------

\section{Introduction}

The time that it takes for an electron to abandon its parent ion when excited
into the continuum by light has been a long-debated issue almost since
the discovery of the photoelectric effect in the early days of quantum
mechanics, when the ``tunnelling'' time problem was first
examined~\cite{maccoll1932}. This question is even difficult to
pose~\cite{landauer1994} -- the deepest underlying problem is probably that
time is not a quantum mechanical operator. In consequence, different
definitions of time have been proposed, and the theoretical discussion lingers
even today~\cite{landsman2015}.

In fact, the topic has been enlivened in recent years by the advances in
attosecond science~\cite{krausz2009,kling2007,scrinzi2006}, and the appearance
of attosecond metrology~\cite{krausz2014}. These developments have enabled to
look in real time into the ionization process, thereby shedding light on the
previously theoretical-only considerations~\cite{pazourek2015}. 
Attosecond streaking~\cite{kitzler2002,itatani2002} is the basic technique
behind the chronoscopy of ionization: a weak attosecond pulse ionizes the
target, and an overlaying longer and intense pulse accelerates the produced
electron, that acquires a final momentum that will depend on the driving field
and on when it was excited into the continuum. 

A particularly successfull attosecond streaking setup is the
attoclock~\cite{eckle2008,pfeiffer2011,pfeiffer2013,landsman2014,landsman2015},
in which the time reference is given by a close-to-circularly polarized laser
field. With this tool, it was possible to measure the ``tunneling delay time''
(interval between the maximum of the electric field and the maximum of the
ionization rate, which was found to be zero within experimental uncertainty),
or the ``electron release time'' (or simply ``ionization time''), a concept
based on a semiclassical picture, defined as the time when the classical
electron trajectory starts in the continuum.
Those are but some of the various \emph{times} that can be defined around the
ionization process. Theoretically, some of the most invoked concepts are perhaps the
Keldysh time~\cite{keldysh1965}, the Buttiker-Landauer
``traversal'' time~\cite{buttiker1982}, or the Eisenbud-Wigner-Smith time-delay
operator~\cite{eisenbud1948,wigner1955,smith1960}.

In this work, however, we adopt a pragmatic, operational definition of
``ionization time'', based on the time at which the occupation of the
continuum states surpasses a certain threshold (close to one). This
can be seen as the time required to ionize completely an ensemble of
identical atoms. This definition is theoretically intuitive, and is
quite useful for our purpose, which is to analyze the possible
variation of the ionization time as a function of the length and shape
of the light pulse. In particular, we employ an optimization algorithm
to find the shape that induces the fastest possible ionization. Or,
put differently, we investigate what is the minimum time required to
ionize one electron for a given laser fluence.  We are particularly
interested in regimes where there are competing mechanisms for
ionization, and on the influence of the details of the electronic
system on the process. The interest is not specifically on the
characteristics of the optimal laser pulse, but on the physical
mechanisms that lead to the fastest possible ionization.

The optimization of quantum processes can be studied theoretically by
quantum optimal control theory~\cite{brif2010,werschnik2007}. Within
this framework, it is also possible to formalize problems where the
target is the duration of the
process~\cite{carlini2007,khaneja2001,moore2012}. In fact, these
methods have already been used to study the optimization of ionization
processes in, e.g., Refs.~\cite{castro2009,hellgren2013}. Here,
however, we are dealing with a multi-objective problem: we try to
determine the shape of the laser field that maximizes ionization
\emph{and} minimizes the total time required for the ionization for a
fixed laser fluence.

There are two different ways to tackle these multi-objective
problems. The simplest one is to write down a single target function,
as a weighted sum of both objectives. The weights, fixed a priori,
determine how relevant is one target versus the other one.  Clearly,
that weighting decision introduces a bias into the objective
functional and therefore into the solution. In optimization theory,
however, there is a fully un-biased procedure to tackle with
multi-objective targets: the Pareto
optimization~\cite{censor1977,bonacina2007}. This will be described in
the next Sec.~\ref{sec:method}, along with the description of the our
model system, and the underlying optimization scheme used to construct
the Pareto front (a differential evolutionary
algorithm). Section~\ref{sec:results} will describe the main
findings. Atomic units, i.e. $e=\hbar=m_e=1$, will be used throughout,
unless stated otherwise.

\section{Method}
\label{sec:method}

We have chosen to work on a model system, defined by the
P\"oschl-Teller potential~\cite{poschl1933}:
\begin{align} \label{eq:poeschl-teller}
 \upt (x) =  -\alpha_m \frac{(2m+1)^2-1}{8\cosh^2{[\sqrt{\alpha_m}x]}}\,,
\end{align}
with $\alpha_m = 2 |E_{\text{GS}}|/m^2$, where $E_{\text{GS}}$ is the
ground state energy, for some integer $m$.  This model has previously
been used to study strong-field photo-ionization~\cite{boucke1997,
  wassaf2003, moiseyev:91}, because it allows to set a fixed number of
bound states (BS) -- given by the integer $m$, and it is regularly
behaved at the origin. The energy of the $n$-th bound state is given
by:
\begin{equation}
E_n^m=-|E_{\text{GS}}|(m-n)^2/m^2\,,
\end{equation}
for $n \in \{0,1,\dots, m-1\}$~\cite{infeld1951}. In this work we will consider
both the case with two bound states, and with only one. We will set,
in analogy with the hydrogen atom, $E_{\rm GS} = -0.5$ Ha.

We will consider the evolution of these systems when irradiated with laser
pulses of duration $T$; as main observable, we use the probability of
ionization at time $T$, when the laser is switched off, as:
\begin{align} \label{eq:ioniz_prob}
 \mathcal{I}(T) = 1 - \sum_{n=0}^{m-1} \bigg| \bracket{\psi_n}{\psi(T)} \bigg|^2,
\end{align}
where $\psi_n$ is the $n$-th bound state, and $\psi(T)$ the evolving
state at time $T$. The physical meaning of this equation is clear: we
measure the occupation of the continuum states, by substracting from
one all occupations of the bound states.

The system is propagated according to the time-dependent
Schr{\"{o}}dinger's equation:
\begin{align} 
\label{eq:tdse}
 i \frac{\partial}{\partial t} \psi(x, t) = 
\bigg[-\frac{1}{2} \frac{\partial^2}{\partial x^2} + \upt(x)  - \epsilon(t) x \bigg] \psi(x,t)\,,
\end{align}
starting from the ground state. The function $\epsilon(t)$ is the
amplitude of the electric field of the laser pulse. Here, we used
the length gauge, and assumed the dipole approximation, neglecting
magnetic effects and considering a space homogeneous electric field.

The electric field $\epsilon(t)$ must be given some functional form,
determined by a set of parameters that define the optimization search
space. In our case, we first set the following form for the vector
potential:
\begin{multline} \label{eq:vect_pot}
 A(t) = -c \epsilon_0 \exp{\bigg[-\frac{(t-T/2)^2}{2\sigma^2}\bigg]} \times  \\ 
        \left\{  \sin{(\omega t + \phi)}/\omega + \sum_{n=2}^{N} d_n \sin{(n \omega t + \phi)}/(n \omega) \right\}.
\end{multline}
from which the electric field is obtained as:
\begin{equation} 
\label{eq:electric_field}
 \epsilon(t) = -\frac{1}{c} \frac{\partial}{\partial t} A(t)\,.
\end{equation}
In these equations: $c$ is the speed of light in vacuum;
$\epsilon_0$ is a parameter that fixes the overall pulse amplitude;
$\omega$ is a base frequency; $\phi$ is a carrier-envelope-phase
parameter that allows to rigidly shift the pulse phase within its
envelope; $d_n$ ($n=2,\dots, N$) form a set of coefficients for the
Fourier expansion in the multiples of $\omega$; $\sigma = T/8$ is the
width of a Gaussian ``envelope'', which is centered at $t=T/2$.  This
Gaussian envelope guarantees that $\epsilon(0) \approx \epsilon(T)
\approx 0$, and a smooth increase of the amplitude. Likewise, the
fact that $A(0) \approx A(T) \approx 0$ ensures that the electric field
integrates to zero $\int_0^T\!\!{\rm d}t\; \epsilon(t) \approx 0$.

The free parameters of the laser, to be explored in the optimization process,
are: its duration $T$, its base frequency $\omega$, the carrier-envelope phase
$\phi$ that determines the position of the field maximum with respect to the
maximum of the envelope, and the coefficients $d_n$ of a Fourier expansion in
the harmonics of $\omega$ up to the $N$th order (in the results presented
below, we used 9 coefficients). These
parameters have to be chosen such that the system is maximally ionized within
minimum laser duration $T$, using a fixed amount of energy, obtained by
integrating its intensity per area:
\begin{align}\label{eq:laser_constraint}
 \tilde{I}= \frac{c}{8\pi}\int_0^T |\epsilon(t)|^2 dt.  
\end{align}
This condition is enforced by adjusting the $\epsilon_0$ parameter, which is
not free, but fixed once all the others are set.  Therefore, the optimizations
are carried out within a space of pulses that carry equal energy per unit area
$a$, $\elaser = a \tilde{I}$ \footnote{
Of course, the parametrization of the electric field amplitude is
just one choice among many possible others; we checked that our results are
qualitatively independent of the parametrization of the electric field by
employing an alternative parametrization, described in
Ref.~\cite{krieger2011}, that respects the same physical constraints.
}.

The time-dependent Schr\"odinger equation was solved with the
Crank-Nicolson propagator, as implemented in the freely available
real-space software suite
\texttt{octopus}~\cite{marques2003,castro2006,andrade2012}. The time
steps were chosen between $0.002$~a.u. and $0.05$~a.u., depending in
each case on the maximal electric field amplitude. The wave function
is discretized in a real-space grid of 0.6~a.u. spacing, enclosed in
large simulation box ($x \in [-110, 110]$ a.u.), enough to exclude
spurious boundary effects.

We now describe the optimization technique, based on the combination
of the Pareto multi-objective process, and the differential evolution
algorithm of Storn and Price~\cite{storn1997}. The optimization search
space is formed, as discussed above, by the $D=N+2$ parameters $T,
\omega, \phi, \lbrace d_n\rbrace_{n=2}^N$, that we group into a
$D$-dimensional vector $p \equiv (p_1,\dots,p_D).$ Each vector or
parameter set determines a laser pulse; the algorithm does not proceed
by iterating or improving one single vector, but many of them. The
success of each pulse is determined by propagating the Schr\"odinger
equation with it, and evaluating the objectives $T$ and
$\mathcal{I}(T)$ (evidently, $T$ does not need to be computed, as it
is directly given by the choice of laser parameters).

Given a set of pulses and their associated objective values, one may then
construct the so-called ``Pareto front''. It is based on the concept of
\emph{dominance}: A pulse is said to be dominated by another laser if it does
not perform better in any of the objectives and performs strictly worse in at
least one. The non-dominated pulses form the Pareto front in the
objective space, spanned by the ionization probability $\mathcal{I}(T)$ and
the laser duration $T$.

The task of finding the Pareto front, by iterating sets of pulse
parameters, is undertaken by the differential evolution algorithm,
which belongs to the genetic family: it starts from a pool of trial
pulses, and \emph{evolves} it, improving their performance with
respect to the objectives, according to rules inspired on biological
evolution~\cite{fraser1957,mitchell1996}. The process starts with a
set of random parent pulses (represented by their corresponding
parameter vectors $p$). Their objective function value is computed,
and then a new generation of pulses is created, by using
\emph{mutation} (i.e. random variation of the parameters) and
\emph{crossover} (i.e. exchange of parameters between desirable
pulses). The evolution algorithm decides how to perform those
operations and generate a new offspring of pulses~\footnote{ The
  reader may consult the original work of Storn and Rice for
  details~\cite{storn1997}; here we merely report the values we used
  for the various adjustable parameters of the process: we set $NP=40$
  for the number of parent pulses that are evolved; $CR=0.9$ for the
  crossover parameter, and used the so-called \emph{dither} technique
  to randomly determine the $F$ parameter as a uniform random number
  within the interval $[0.5,1]$.  }

Once a generation and its offspring has been generated, the \emph{selection}
must be performed to choose the survivors, in order to proceed with the
algorihm. We have adopted a two-fold selection
strategy~\cite{babu2005,mezura2008}: First, on the individual level, a trial
pulse with more desirable values of ionization and laser duration than its
direct predecessor substitutes it. Second, on a global level,
the principle of \textit{dominance} is applied: Dominated
lasers are excluded from the pool of successful lasers and consequently
re-initialized before being allowed to spawn again. The nondominated lasers
form the Pareto front in the objective space spanned by ionization and
laser duration. Together with the set of re-initialized random lasers the Pareto front
lasers represent the next generation.

A convergence criterion must be set to finish the process: in the
calculations shown below, the process is stopped when either the
Pareto front stabilizes for 200 consecutive iterations, or
after a maximum number of 10\,000 differential evolution
iterations. The procedure is repeated for each laser energy $\elaser$
several times to rule out a bias due to the initial random
configuration.

\section{Results}
\label{sec:results}

\begin{figure}
 \begin{center}
  \includegraphics[width=0.99\columnwidth,clip]{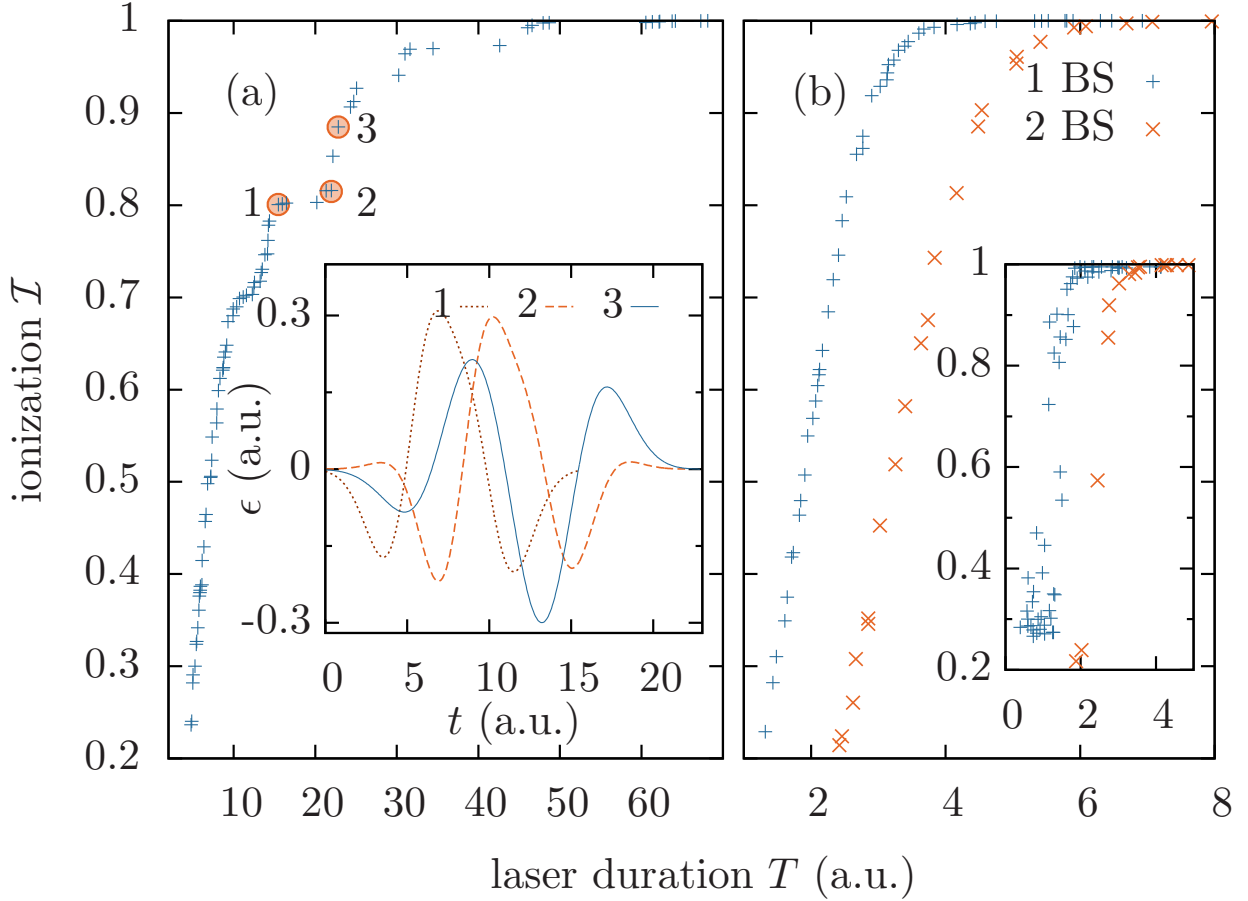} 
  \caption{\label{fig:pareto_step} (Color online) (a) Ionization of
    the system with one bound electron as a function of laser duration
    for the optimal lasers with $\elaser=2$ Ha. The inset shows the
    lasers amplitudes $\epsilon(t)$ for the marked points.  (b)
    Ionization at $\elaser=60$~Ha., for the system with only one bound
    state (blue), and with two bound states (red).  The inset shows
    $\mathcal{I}$ as a function $T$ for the same lasers using a
    classical model.}
 \end{center}
\end{figure}

Figure~\ref{fig:pareto_step} shows the Pareto fronts for two sample
laser energies: $\elaser=2$~Ha (a) and $\elaser=60$~Ha (b). For fixed
laser fluence, the total ionization increases monotonically with the
total duration of the laser. Also, as expected, more energetic lasers
can ionize faster the electron.

In the left panel (a) the front is not smooth, exhibiting a series of
steps. They are observable up to $\elaser=10$~Ha. The inset proves how
these steps are due to a qualitative change in the laser shapes, such
as the addition of one half-cycle by which the lasers labeled 2 and 3
differ. Such abrupt changes in the ionization behaviour, due to
subcycle dynamics, has also been observed
experimentally~\cite{uiberacker2007} in the regime of nonadiabatic
tunneling~\footnote{ The intermediate regime between the purely
  adiabatic tunneling ionization, and the multiphoton ionization, has
  been termed nonadiabatic tunneling.  }. The steps can be
alternatively understood in terms of changes of the carrier envelope
phase $\phi$, and this fact underlines the importance of this phase in
this ionization regime, as lasers 2 and 3 have essentially the same
duration $T \approx 23$~a.u. and the same frequency $\omega \approx
0.67$~Ha, differing only by $\phi$.

%% Such steps, due to subcycle dynamics, have been observed also
%% experimentally~\cite{uiberacker:07} in the regime of nonadiabatic
%% tunneling. To put it alternatively in terms of the carrier-envelope phase
%% $\phi$, and not the numbers of half-cycles, this result underlines the
%% importance of $\phi$ in this ionization regime~\cite{chelkowski:05, paulus:01,
%%   yudin:01} as lasers 2 and 3 have essentially the same duration $T \approx
%% 23$~a.u. and the same frequency $\omega \approx 0.67$~Ha.

For higher laser energies, as in the case shown in the right panel
(b), a complete ionization is obtained in one cycle, and the Pareto
front becomes smoother. For $\elaser=60$~Ha a simple classical model
even suggests over-the-barrier ionization instead of nonadiabatic
tunneling ionization: the laser deforms the atomic field $\upt$, such
that the electron density is spilled out. Using a simple classical
model, we initialized a distribution of point-like, noninteracting
classical particles according to the ground-state density of the
potential given by Eq.~\eqref{eq:poeschl-teller}. By following their
trajectories in the combined potential $\upt(x)-\epsilon(t)x$, we
could determine if, and when, they cross this barrier. Despite its
simplicity (most notably, the quantized bound states are missing),
this model reproduces well the differences in ionization times for the
two different $\upt$ (with 1 BS and 2 BS): $\sim \!2$~a.u. (cf. the
inset in Fig.~\ref{fig:pareto_step}b).

\begin{figure}
 \begin{center}
  \includegraphics[width=0.99\columnwidth,clip]{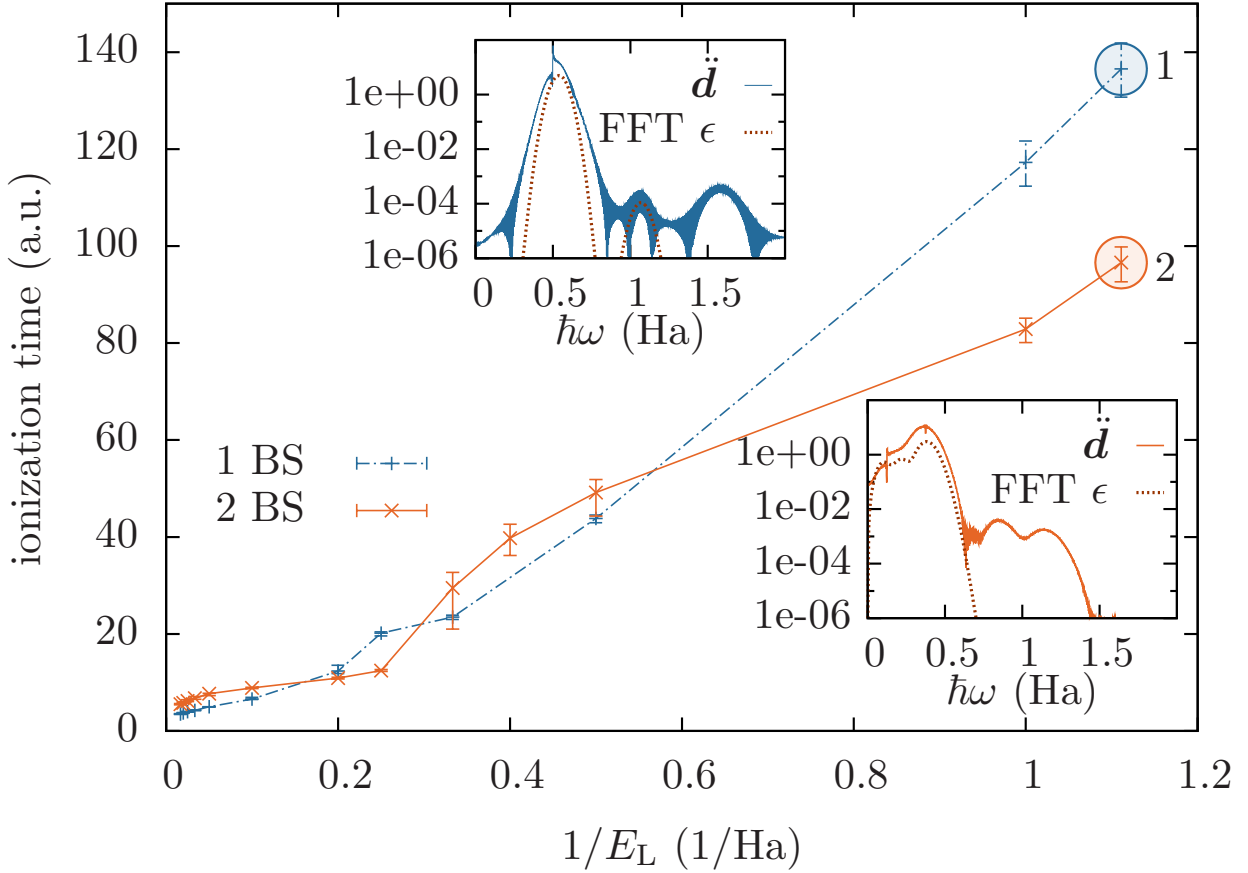} 
  \caption{\label{fig:ene_vs_mintime} (Color online) Ionization time
    as a function of inverse laser energy. The insets show the dipole
    acceleration of the electron system, and the Fourier transform
    (FFT) of the laser at the marked points of the main panel (top
    inset for point 1, bottom inset for point 2).}
 \end{center}
\end{figure}

The analysis of the Pareto fronts permits to define a total
``ionization time'' for each laser energy $\elaser$ by intersecting
the front at a given ionization probability. In the following we will
take this threshold to be $\mathcal{I}(T)=0.98\pm 0.005$, i.e., the
almost complete ionization of the electron.  In
Fig.~\ref{fig:ene_vs_mintime} we report the dependence of these total
ionization times on the inverse laser energy for the 1 BS system and
the 2 BS system (the presence of the error bars is due to the use of
an uncertainty of $\pm 0.005$ in the above definition).  The range of
$1/\elaser$ covers all ionization regimes: the above mentioned
over-the-barrier ionization, nonadiabatic tunneling and multiphoton
ionization. Note that the average intensity of the lasers under
consideration in SI units is between $3 \times 10^{13}$ and $10^{17}$
W/cm$^2$.  We prefer not to make use of cycle-averaged quantities, as
the Keldysh parameter~\cite{keldysh1964}, to distinguish the regimes,
due to the very short duration of the pulses.

At high laser energies (left part of the graph in
Fig.~\ref{fig:ene_vs_mintime}), one full cycle is sufficient to ionize
both the 1 BS and the 2 BS systems; the 1 BS system ionizes first. The
situation is changed at around $1/\elaser=0.25$~Ha$^{-1}$: at those
energies, the second bound state of the 2 BS system serves as an
intermediate state for the electron, before leaving the atom -- the
so-called shake-up-and-ionize mechanism~\cite{uiberacker2007}. As a
consequence, for the 2 BS system a single full cycle suffices to
ionize the atom, while for the 1 BS system the shortest ionizing laser
consists of three half-cycles. The situation is reversed between
$1/\elaser = 0.4$ and 0.6~Ha$^{-1}$, but the shake-up-and-ionize
mechanism produces an even faster ionization at the lowest energies
(highest inverse \elaser).  In fact, at those energies, the process
for the 2 BS system is better understood by considering it in the
nonadiabatic tunneling regime, while for the 1 BS system the process
is best understood in the multiphoton absorption regime.

The insets in Fig.~\ref{fig:ene_vs_mintime} may help to understand
this. They show the Fourier transform (in fact, the power spectrum) of
the field and of the acceleration of the dipole moment, for the 1 BS
system (top) and for the 2 BS system (bottom), for the circled points,
tagged 1 and 2 respectively. According to the dipole acceleration
$\ddot{\bm{d}}$, i.e. the optical absorption/emission of the electron
system, the 1 BS system mainly absorbs at the ionization potential
$0.5$~Ha, while the 2 BS system absorbs at lower energies, i.e. at the
frequency of the bound-bound transition ($0.38$~Ha), and at the
ionization energy from the excited-state ($0.12$~Ha). Consequently,
at low laser energy, the electron is excited to the excited-state
(shake-up), and subsequently tunnels out. This route is faster than
single- or multiphoton absorption of the 1 BS system.

The upper inset also shows the power spectrum of the laser, for the 1
BS system.  Indeed, the first two peaks ($0.5$ and $1.1$~Ha) of the
power spectrum of the dipole acceleration are also present in the
power spectrum of the laser. The third peak at $1.6$~Ha in the power
spectrum of the dipole acceleration can then be interpreted as the
absorption of two distinct photons of the former energies.  In the
lower inset, corresponding to the 2 BS system, one observes, apart
from the excitation energies ($0.12$~Ha and $0.38$~Ha) in the Fourier
transform of the laser and in the absorption/emission, two further
peaks for $\ddot{\bm{d}}$ at $\sim \!0.84$~Ha, the $7$th harmonic of
$0.12$~Ha, and at $\sim \! 1.14$~Ha, the $3$rd harmonic of $0.38$. The
generation of high harmonics is best understood in terms of tunneling,
by making use of the three-step model~\cite{corkum1993}.

\begin{figure}
 \begin{center}
  \includegraphics[width=0.99\columnwidth,clip]{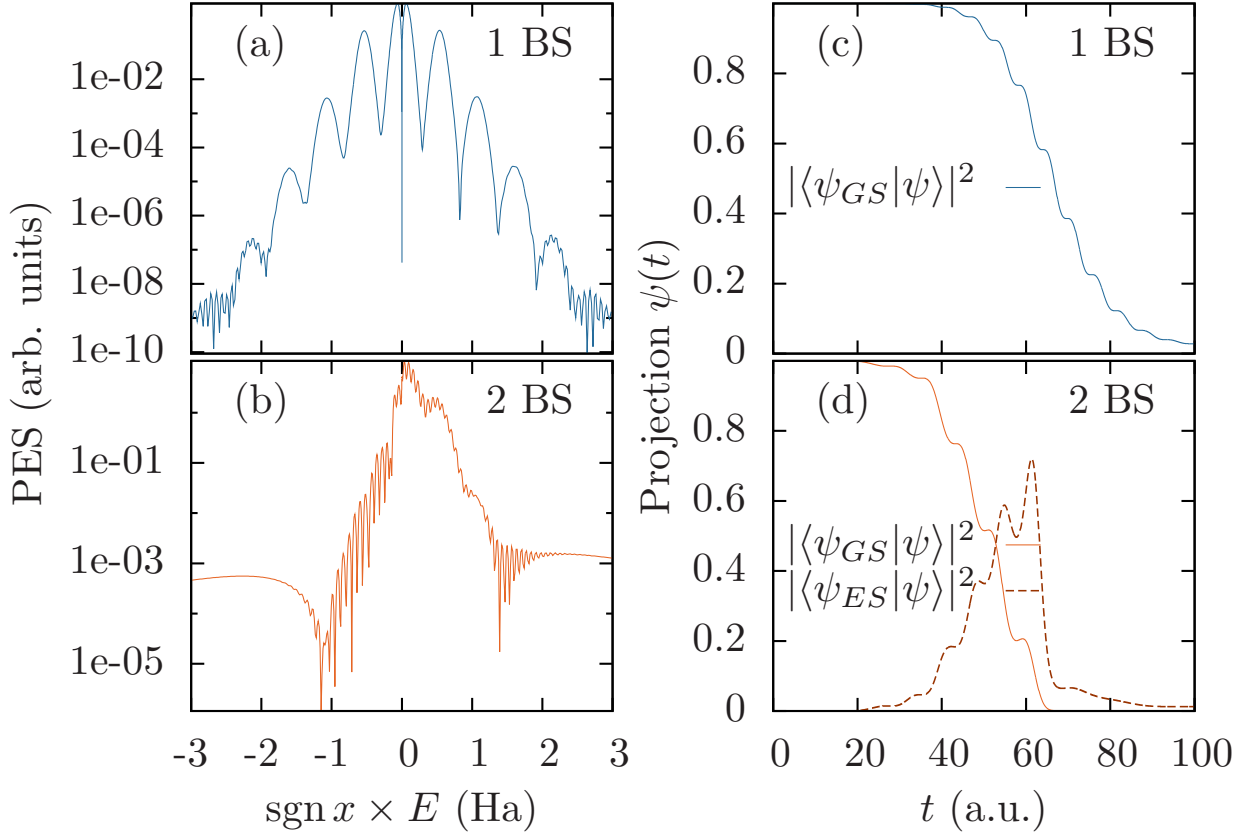} 
  \caption{\label{fig:pes_proj}
    (Color online) Angle-resolved photoelectron spectra for the 1 BS system
    (a) and the 2 BS system (b), 
     and projections of the evolving wavefunction for 1 BS (c) and 2 BS (d) at $1/\elaser=1.1$~Ha.
%     The peaks of the spectrum for 1 BS are distanced by $0.5$~Ha, the single photon energy, which is characteristic for multiphoton absorption. 
%     For 2 BS the absence of a clearly structured and asymmetric spectrum hints at tunneling ionization. Indeed, panel (d)
%     shows a sequential excitation from the GS to the ES from where the
%     electron escapes by tunneling. 
    } 
 \end{center}
\end{figure}
   
Figure~\ref{fig:pes_proj} further analyzes the ionization dynamics at
$1/\elaser=1.1$, the lowest laser energy, illustrating again how the two
different ionization mechanisms of the two systems (with and without one bound
excited state) produce different spectroscopic signatures. Panels (a) and (b)
show the ``angle-resolved'' (in 1D: left-right) photoelectron spectrum of both
systems. The spectrum of the 1 BS system exhibits equally distanced peaks with
decreasing intensity. This spectrum shape is typical of above threshold
ionization, where the electron absorbs more photons than necessary to
overcome the ionization barrier. The symmetry is due to the fact that the
ionization is considered to be ``vertical'', i.e. independent of any preferred
direction of the field. The spectrum of the system with 2 BS, instead, differs
in two points: first, it is not symmetric with respect to left and right from
the atomic position, second, it does not have distinct, equally distant
peaks. Both properties are characteristics typical of nonadiabtaic
tunneling~\cite{chelkowski2005}.

Finally, Fig.~\ref{fig:pes_proj} displays in panels (c) and (d) the
projection of the propagated electron wavefunction onto bound
field-free states. It once again helps to understand how the two
ionizations proceed differently. The electron of the 2 BS system is
first promoted to the excited state, leading to a peak in the
projection of the wavefunction on it [Fig.~\ref{fig:pes_proj}(d)], and
then tunnels out of the atomic potential.

%---------------------------------------------
% CONCLUSION
%---------------------------------------------

\section{Conclusions}

We have analyzed the time it takes to photo-ionize a hydrogen-like model
system, and how the variation of the shape of the laser pulse may be
used to accelerate the process. For this purpose, we set-up a
Pareto optimization scheme with the double objective of increasing
ionization and reducing ionization time, and used a genetic-type
algorithm (the differential evolution) to perform the
optimizations. The search was performed on spaces of pulses carrying
fixed energies per unit area, and we looked at different regimes
by considering differents energies. The presence or absence of
intermediate bound states was found to be relevant, since it may
determine what is the fastest ionization channel. The
shake-up-and-ionize mechanism may in fact lead to the fastest channel,
depending on the energy carried by the laser pulse. The process may
also have signatures of tunneling, or of multi-photon ionization. The
possibility of designing pulse shapes that significantly accelerate
ionization may be relevant to shed light into the problem of defining
and measuring the times of these processes, and may be of use for the
experimental design of atto-second resolution chronometers.

%% by taking into account i) the number of excited states and ii) constraints on
%% the laser energy. To this end we used a Pareto optimization whose set of
%% optimal solutions provided insight into subcycle dynamics. The presence of a
%% second bound state reduces the energy of main absorption which is advantageous
%% at low laser energies, as it allows sequential ionization. Once the laser
%% energy is high enough for ionization within one laser cycle, the system with
%% the higher absorption energy, i.e. the system with only one bound state, is
%% ionized faster.

%---------------------------------------------------
%  BIBLIO AND STUFF
%---------------------------------------------------
\begin{acknowledgments}
D.~K. and M.~A.~L.~M. acknowledge financial support from the French ANR
(ANR-08-CEXC8-008-01). D.~K. was also financed by the Joseph Fourier
university funding program for research (p\^ole Smingue). Computational
resources were provided by GENCI (project x2011096017). D.~K. is indebted to
Lauri Leehtovaara for animating discussions. AC acknowledges support from the
Spanish Grants No. FIS2013-46159-C3-2P and FIS2014-61301-EXP.
\end{acknowledgments}

% using kbibtex
\bibliography{maxi_ionization.bib}

\end{document}